\documentclass[useAMS,usenatbib]{mn2e}
\usepackage{graphicx}
\newcommand {\be}{\begin{equation}} 
\newcommand{\ee}{\end{equation}}    
\newcommand{\sss}{\scriptscriptstyle}

\def\ddt{\frac{\partial}{\partial t}}

\def\vti{v_{{\sss T}i}}

\def\vtj{v_{{\sss T}j}}

\title[Electric fields in solar magnetic structures  ]{
Electric fields in solar magnetic structures due to gradient driven
instabilities: heating and acceleration of particles}
\author[J. Vranjes and S. Poedts]{J. Vranjes\thanks{E-mail:
Jovo.Vranjes@wis.kuleuven.be; jvranjes@yahoo.com} and S.
Poedts\thanks{E-mail:
Stefaan.Poedts@wis.kuleuven.be}\\
K. U. Leuven, Center for Plasma Astrophysics, Celestijnenlaan 200B,
3001 Leuven,
 Belgium,\\ and Leuven Mathematical Modeling and Computational Science Center
 (LMCC)}
\begin{document}

\date{Accepted xxx. Received xxx; in original form xxx}

\pagerange{\pageref{firstpage}--\pageref{lastpage}} \pubyear{2002}

\maketitle

\label{firstpage}

\begin{abstract}
The electrostatic instabilities driven by the gradients of the density, temperature and magnetic field, are discussed
in their application to solar magnetic structures. Strongly growing modes are found for some typical plasma parameters.
These instabilities  i)~imply the presence of  electric fields that can accelerate the plasma particles in both
perpendicular and parallel directions with respect to the magnetic field vector, and ii)~can stochastically heat ions.
The perpendicular acceleration is to the leading order determined by the $\bmath{E}\times \bmath{ B}$-drift acting equally on
both ions and electrons, while the parallel acceleration is most effective on electrons.  The experimentally confirmed stochastic
heating is shown to act mainly in the direction perpendicular to the magnetic field vector and acts stronger on heavier ions. The energy release rate and heating
may exceed for several orders of magnitude the value accepted as necessary for a self-sustained heating in the solar corona. The energy source for both
the acceleration and the heating is stored in the mentioned background gradients.
\end{abstract}


\section{Introduction}

Very strong electric fields have been reported in the solar atmosphere in many studies in
the past. The magnitude of such fields may amount to $7\cdot 10^4\;$V/m \citep{dav} and
even up to $1.3 \cdot 10^5\;$V/m \citep{zs}. It is widely believed that these electric fields appear in the process of magnetic reconnection. The magnetic reconnection itself implies a considerable change
in the magnetic field topology, and it is also used in describing rapid energy releases and
`energization' of plasma particles, i.e.\ their heating and acceleration. However, at
least in some cases \citep{jan,mc,pud} such energy release events seem to appear without
a measurable change of the  magnetic energy and configuration and, hence, a different
description and source of such events may be needed. In the present work, we show that such an
alternative source naturally exists, and that it may drive extremely strong electrostatic
instabilities and produce electric fields of the order mentioned above. The energy for
such instabilities is provided by the omnipresent gradients of the background plasma
parameters (the density, temperature, and magnetic field). However, the physics and
the features of this source, and the mechanism by which its energy is transferred into these
instabilities are beyond the standard magnetohydrodynamics (MHD) model. In order to describe
it, a multi-component (fluid or kinetic) theory is needed.

Typical studies of oscillations and instabilities in the solar
magnetic field structures imply a cylindric geometry with parameters
having different values inside and outside of the structure, yet
having at the same time some constant values in the two separate
domains. Moreover, the usually curved structures (e.g.\ magnetic
loops) are often flattened out in the modeling and the curvature effects
are thus neglected in the simplified models.
In reality however, the plasma parameters (the density,
temperature and magnetic field magnitude) may vary both in the axial
(parallel to the magnetic field vector) and in the radial
(perpendicular) direction. The axial variation is typically on a
much larger characteristic spatial scale  in comparison to the
radial variation, and in some cases can be omitted. The presence of
these radial inhomogeneities in the background of some accidental
fluctuations implies a source of energy for certain types of
instabilities. These background (equilibrium) gradients also imply
the equilibrium drift velocities, and the associated instabilities
are usually termed as reactive drift  instabilities, first predicted
long ago by \citet{rs1}. A drift wave driven by the density gradient
is typically growing due to the electron thermal effects in both
collisional and collision-less regimes \citep{v1,v2,v3}. The former
is well described within the two-component fluid theory, the latter
however, is a strictly kinetic effect. In both cases the ions play
a stabilizing role, and in some situations they may even impose a
threshold for the instability.

However, in the case of hot ions and in the presence of both the density and
temperature gradients, the above mentioned reactive instability is
termed as $\eta_i$-instability, where now the ions play a crucial
destabilizing role. Here, $\eta_i=L_n/L_{\sss T}$, and $L_n=(d
n_0/dx/n_0)^{-1}$, $L_{\sss T}=(d T_0/dx/T_0)^{-1}$ are the
characteristic inhomogeneity scale-lengths of the equilibrium
quantities that are here, and further in the text, denoted by the
index 0. The coordinate $x$ in the present local analysis is used to
describe the changes in the radial (perpendicular) direction. The
background magnetic field is typically also inhomogeneous (i.e., with a
curvature and a gradient in the perpendicular direction), and this
may be described by yet another characteristic scale-length $L_{\sss
B}=(d B_0/dx/B_0)^{-1}$. The interplay of these three gradients
determines the behavior of low frequency ($\omega \ll \Omega_i= q_i
B_0/m_i$, $q_i=Z_i e$ is the ion charge) electrostatic oscillations
and instabilities. This will be demonstrated in the forthcoming
text.

\section{Electrostatic instability in an advanced fluid model}

We apply the two-fluid model developed in numerous works related to
laboratory plasmas \citep{w1,w2}. A systematic presentation of the
theory that has been successfully used in the past in the prediction
of transport processes in tokamak plasmas can be found in
\citet{w3}. There, one can also see a complete agreement between
this advanced fluid model and the kinetic theory (that is one of the reasons
for the term {\em `advanced'} used here). Part of the basic theory of the
drift wave applied to the solar plasmas is given in
\citet{v1}. It has also been used very recently \citep{v2,v3} in
order to explain some essential properties of the coronal heating mechanism.
The theory is described in detail in the references mentioned above.
For completeness, we shall provide here  a general description
of the derivations, emphasizing some most important features of the
model and providing explanations for the assumptions used in the
procedure. The present analysis is restricted to the electrostatic
limit. Note however, that the theory works well also in the full
electromagnetic limit \citep{and}, where it can be used e.g.\ for
describing the ballooning instabilities. This domain also can be of
great importance for the solar plasma as it may provide a triggering
mechanism for abrupt changes in the magnetic field topology, e.g.\ in processes like
magnetic reconnection and Coronal Mass Ejections (CMEs).

The presence of hot ions (typical for the solar atmosphere) and the
background temperature gradient (that is expected in solar magnetic
configurations) implies, first of all, the necessity of including their
full thermal response (the pressure and the gyro-viscosity
collision-less stress tensor) in the momentum equation, and, second,
the use of the ion energy equation in the mathematical model.
For the present purpose, the
later comprises the diamagnetic heat flow term  only, and can be
written as \citep{w3}
\be
\frac{3}{2} n_i \left(\ddt + \bmath{v_i} \cdot \nabla\right) T_i + p_i
\nabla\cdot \bmath{v_i}= - \nabla \cdot\bmath{q_{*i}}, \label{e1} \ee
\[
\bmath{q_{*i}}= \frac{5}{2} \frac{n_i T_i}{m_i \Omega_i} \bmath{e_{\|}}\times \nabla T_i.
\]
Here, $T_i$ is in energy units, $\bmath{ q_{*i}}$ is the diamagnetic
heat flux, and $\bmath{ e_{\|}}= \bmath{ B}/B$. The given  form of $\bmath{
q_{*i}}$ can be obtained directly from the drift-kinetic theory. In
the case of different temperatures (pressures) in the two directions
\citep{m} it is to be replaced with $[p_{i\bot}/(m_i \Omega_i)]\bmath{
e_{\|}}\times \nabla (2 T_{i\bot} + T_{i\|}/2) + [(p_{i\|} -
p_{i\bot})/(m_i \Omega_i)] T_{i\|} \bmath{ e_{\|}}\times (\bmath{ e_{\|}}
\cdot \nabla) \bmath{e_{\|}}$. A detailed analysis of the temperature
anisotropy effects on the gradient driven instability is performed
by \citet{m2}.

The magnetic field is inhomogeneous in the general case. This implies
that, in the continuity equation, the contribution of the diamagnetic
drift to the ion flux does not vanish, and the  appropriate
linearized term is
\[
\nabla \cdot (n_i \bmath{v_{*i}})= \frac{1}{T_i} \bmath{
v_{bi}}\cdot \nabla p_{i1} \neq 0.
\]
For the same reason  we have also  $\nabla
\cdot \bmath{ v_{\sss E}}\neq 0$, where $\bmath{ v_{\sss E}}$ is the $\bmath{
E}\times \bmath{ B}$-drift. Hence, an additional magnetic drift velocity
$\bmath{ v_{bi}}$ appears in the description of the ion motion. We use
standard notation from the drift wave theory where $\bmath{ v_{*i}}=\bmath{
e_{\|}} \times \nabla p_i/(q_i n_i B)$. In the equations above we
have $\nabla \cdot \bmath{ q_{*i}}= -5 n_i v_{*i} \nabla T_i/2 + 5 n_i
\bmath{ v_{bi}}\cdot \nabla T_i/2$. The first (non-curvature) part in
this expression cancels out in the procedure of calculating $\nabla
\cdot \bmath{ v_i}$ in the ion continuity equation. The second term comprises the
ion magnetic drift, which in the general case is the sum of the
curvature and the gradient-$B$ drifts
\[
\bmath{ v_{bi}}= \frac{v_{\|}^2}{\Omega_i} \bmath{ e_{\|}} \times (\bmath{
e_{\|}} \cdot \nabla) \bmath{ e_{\|}} + \frac{v_{\bot}^2}{2 \Omega_i}
\bmath{ e_{\|}} \times \nabla \log B.
\]
Here, $(\bmath{ e_{\|}} \cdot \nabla) \bmath{ e_{\|}}=- \bmath{ R}/R^2$ and $R$
denotes the radius of the curvature of the magnetic field, while the
two velocities are in general case different $v_{\|}^2= T_i/m_i$,
$v_{\bot}=2 T_i/m_i$. In what follows, we shall use the expression
for the  effective total curvature drift \citep{w3} $\bmath{
v_{bi}}\simeq [2 T_{i0}/(q_i B_0)] \bmath{ e_{\|}}\times (\bmath{e_{\|}}
\cdot\nabla) \bmath{ e_{\|}}$.

\begin{figure}
\includegraphics[height=6cm, bb=15 15 275 222, clip=,width=.95\columnwidth]{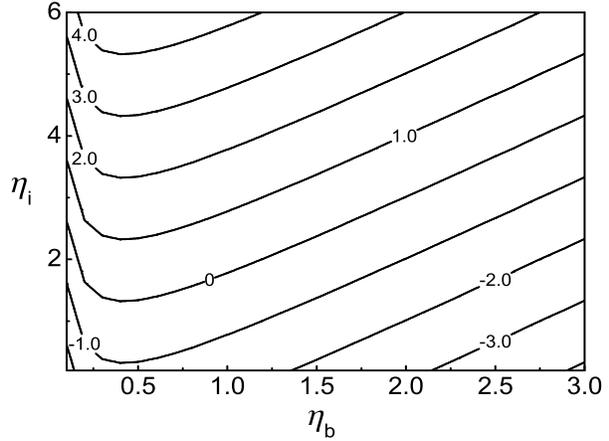}
\vspace*{3mm} \caption{Contour plot of $F(L_n, L_{\sss T}, L_{\sss
B})=\eta_i - \eta_{th}$, for $k_y \rho_s =0.1$. The positive values
correspond to the instability. }\label{fig1}
\end{figure}

The perturbed ion perpendicular velocity can be obtained from the
ion momentum equation by applying the vector product $\bmath{ e_\|}
\times ...$, yielding
\be
 \bmath{ v_{i\bot}}=\bmath{ v_{\sss E}} + \bmath{ v_{*i}} + \bmath{ v_{pi}} +
\bmath{ v_{\pi i}}. \label{vel} \ee
Here, $\bmath{ v_{\sss E}}$ and $ \bmath{ v_{*i}}$ are already defined above,
the  third  term  is the ion
polarization drift,  and $\bmath{ v_{\pi i}}$ denotes the drift due to the
ion stress tensor effects. For small accidental perturbations
propagating predominantly in the perpendicular direction $\sim
\exp[- i \omega t + i k_y y + i k_z z]$, $|k_y| \gg |k_z|$,
$|\omega| \ll \Omega_i$,  the linearized ion energy equation yields
\citep{w3}
\[
\frac{T_{i1}}{T_{i0}} = \frac{\omega}{\omega- 5 \omega_{bi}/3}
\left[\frac{2}{3} \frac{n_{i1}}{n_0} +
\frac{\omega_{*e}}{\omega}\left(\eta_i - \frac{2}{3}\right) \frac{e
\phi_1}{T_{e0}}\right].
\]
Here, the terms $\omega_{*j}$, $\omega_{bj}$ are the product of the
perpendicular wave number component $k_y$ and the diamagnetic and
magnetic drifts, respectively. Note that $\omega_{*e}=k_y v_{*e}= -
\tau \omega_{*i}$, $\tau=T_{e0}/T_{i0}$. Using Eq.~(\ref{vel}) in the
 ion continuity, the ion  density perturbation can be written as
 \citep{w3}:
\[
\frac{n_{i1}}{n_0}= \left\{\omega (\omega_{*e}- \omega_{be}) +
\left(\eta_i -\frac{7}{3} + \frac{5 \eta_b}{3}\right) \omega_{*e}
\omega_{bi} \right.
\]
\[ \left.
 - k_y^2 \rho_s^2 \left[\omega - \omega_{*i} (1+
\eta_i)\right] \left(\omega - \frac{5\omega_{bi}}{3}\right)\right\}
\]
\be
 \times
\left[\omega^2 - \frac{10 \omega \omega_{bi}}{3} + \frac{5
\omega_{bi}^2}{3}\right]^{-1} \frac{e \phi_1}{T_{e0}}. \label{e2}
\ee
The electron parallel dynamics yields just the Boltzmann
distribution for the electron number density, and using the
quasi-neutrality condition, one then obtains the dispersion equation
\[
\Omega^2\left(1+ k_y^2 \rho_s^2\right) + \Omega \left[ \frac{10
\eta_b}{3 \tau} + k_y^2 \rho_s^2 \frac{5 \eta_b}{3 \tau} -1 + \eta_b
+ k_y^2 \rho_s^2  \frac{1+ \eta_i}{\tau}\right]
\]
\be
+\frac{5 \eta_b^2}{3 \tau^2} + \left(\eta_i - \frac{7}{3} + \frac{5
\eta_b}{3} \right) \frac{\eta_b}{\tau} + k_y^2 \rho_s^2 \frac{1+
\eta_i}{\tau} \frac{5 \eta_b}{3 \tau} =0. \label{e4} \ee
Here, $\Omega\equiv \omega/\omega_{*e}$, $\rho_s =c_s/\Omega_i$, $c_s^2= T_{e0}/m_i$, $\eta_b=\omega_{bi}/\omega_{*i}=
L_n/L_{\sss B}$, and $\omega_{be}=- \tau \omega_{bi}$.  It is seen that, without the magnetic field inhomogeneity, Eq.~(\ref{e4}) yields only the standard drift mode driven by the density gradient, with  $\omega\sim 1/L_n$. The  magnetic
drift terms $\omega_{bi}$ are responsible for the appearance of the additional plasma mode, while all three gradients together are
responsible for the instability.
 As shown by \citet{w3}, Eq.~(\ref{e4}) can be
solved analytically yielding the approximate growth-rate (normalized to $\omega_{*e}$)
\[
\gamma=\frac{(\eta_b/\tau)^{1/2}}{1+ k_y^2 \rho_s^2} (\eta_i -
\eta_{th})^{1/2},
\]
\[
\eta_{th}= \frac{2}{3} - \frac{\tau}{2} + \eta_b\left(\frac{\tau}{4}
+ \frac{10 }{9 \tau} \right)
 + \frac{\tau}{4 \eta_b}  - \frac{k_y^2\rho_s^2}{2 \eta_b} \left[
 \frac{5}{3}  - \frac{\tau}{4
 \eta_b} \right.
 \]
 \[
 \left.
 + \frac{\tau}{ 4 \eta_b} - \left(\frac{10}{3} +
 \frac{\tau}{4} - \frac{10}{9 \tau} \right) \eta_b +
 \left(\frac{5}{3} + \frac{\tau}{4} - \frac{10}{9 \tau}\right)
 \eta_b^2\right].
 \]

\begin{figure}
\includegraphics[height=6cm, bb=19 15 300 222, clip=,width=.95\columnwidth]{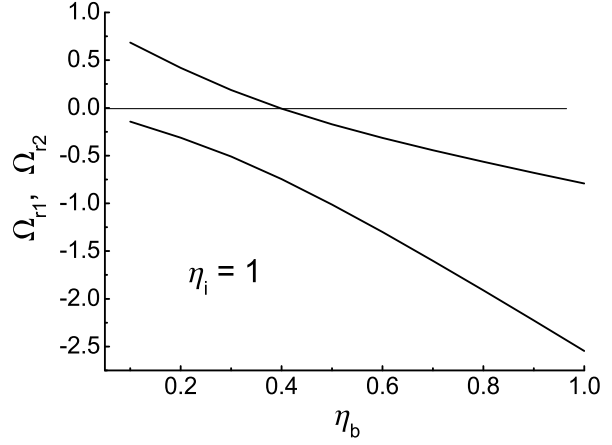}
\vspace*{3mm} \caption{Two real solutions (normalized to
$\omega_{*e}$)  in the case $\eta_i=1$ and $k_y \rho_s =0.1$.
}\label{fig2}
\end{figure}
It is seen that for $\eta_i>\eta_{th}$ there will be an instability. For a fixed $\tau$, the
stability conditions are completely determined by the three
characteristic inhomogeneity scale-lengths $L_n, L_{\sss T}, L_{\sss
B}$. This is demonstrated in Fig.~1 where we give the contour plot
of $F(L_n, L_{\sss T}, L_{\sss B})=\eta_i - \eta_{th}$ for $k_y
\rho_s =0.1$ and for $\tau=T_{e0}/T_{i0}=1$. The positive lines
denote the values for which the gradient-driven instability takes
place. In  application to the solar atmosphere with
$T_{e0}=T_{i0}=10^6\;$K,  and assuming $B_0=3\cdot 10^{-2}\;$T, for
hydrogen ions we have $\rho_s=0.032\;$m and, hence,  the condition
$k_y \rho_s =0.1$ implies the perpendicular wavelength $\lambda_y=2\;$m.
The assumption of the nearly perpendicular perturbed ion motion
for such a short perpendicular wave-length in fact implies a
flute-like mode that is very elongated along the magnetic field
vector, with a parallel wave-length that is measured in hundreds
of kilometers. Such strongly elongated modes (i.e.\ $k_\bot\gg
k_{\|}$)  are easily excited under laboratory conditions (e.g.\ in a
tokamak plasma) in spite of the rather limited scales in the parallel
direction. In the solar magnetic structures with naturally drastic
differences in the perpendicular and parallel scale-lengths, their
excitation  is expected  to be even more efficient.

The second order dispersion equation (\ref{e4}) is solved
numerically and some results are presented in Figs.~2-5. In
accordance with Fig.~1, in Fig.~2 for $\eta_i=1$ we have two real
solutions for the wave frequency, one essentially due to the density gradient and the other due to the magnetic field gradient.
The initially positive solution
changes the sign for $\eta_b\simeq 0.4$ and then both solutions
propagate in the direction of the ion diamagnetic drift. Note that
for the parameters used above and for  $k_y \rho_s =0.1$,  the
normalized frequency $\Omega\sim 1\;$Hz here implies  $L_n=L_{\sss
T} \sim 10^3\;$m.

In Fig.~3, the case $\eta_i=3$ is presented. Here, both  initial solutions are
positive and real  for small values of $\eta_b$. They   merge at
around $\eta_b\simeq 0.09$, yielding a pair of complex-conjugate
solutions with the real part becoming  negative for $\eta_b>0.25$. The
instability vanishes for  $\eta_b>2.04$ when  two negative real
solutions appear. The mode is particularly strongly growing
($|\gamma| >|\Omega_r|$) in the range $\eta_b\in (0.15, \, 0.7)$.

\begin{figure}
\includegraphics[height=6cm, bb=15 15 283 215, clip=,width=.95\columnwidth]{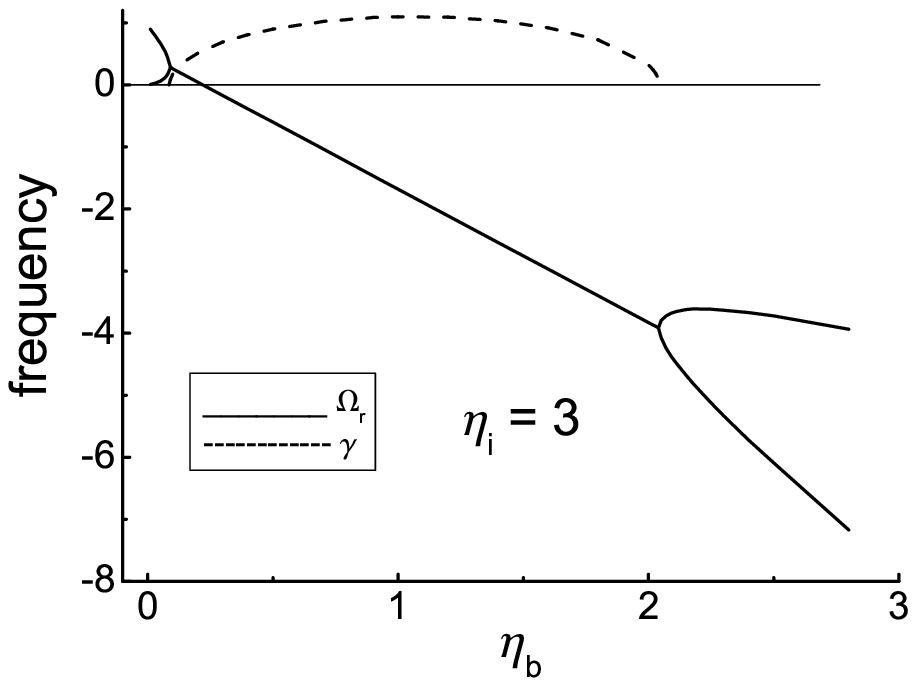}
\vspace*{3mm} \caption{The solutions of Eq.~(\ref{e4}) for
$\eta_i=3$ showing two real solutions for small $\eta_b$ and for
$\eta_b>2.04$. The dashed line is the growth-rate  for the
complex-conjugate solutions in between. }\label{fig3}
\end{figure}

\begin{figure}
\includegraphics[height=6cm, bb=15 15 283 222, clip=,width=.95\columnwidth]{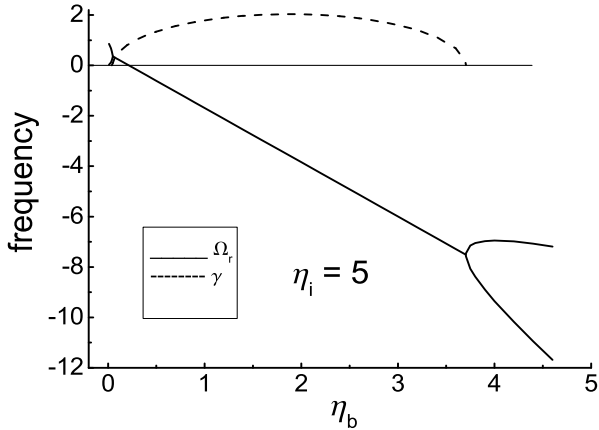}
\vspace*{3mm} \caption{The solutions of Eq.~(\ref{e4}) for
$\eta_i=5$. }\label{fig4}
\end{figure}

\begin{figure}
\includegraphics[height=6cm, bb=15 15 283 222, clip=,width=.95\columnwidth]{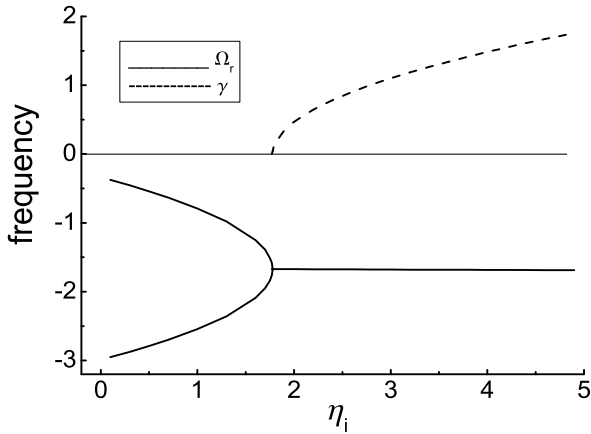}
\vspace*{3mm} \caption{The solutions of Eq.~(\ref{e4}) in terms of
$\eta_i$ for $\eta_b=1$.   }\label{fig5}
\end{figure}

Considering $\eta_i=5$,  in Fig.~4 we present the same mode behavior as
in Fig.~3. Such a larger $\eta_i$ value yields a growth-rate that is larger by
about a factor 2 and the instability range in terms of $\eta_b$ is
widened  to $\eta_b\in (0.05, \, 3.7)$.

In Fig.~5, for a fixed
$\eta_b=1$ and $k_y \rho_s=0.1$, the frequency is calculated in
terms of $\eta_i$. The two negative real solutions (c.f.\ Fig.~2)
merge for $\eta_i\geq 1.78$ yielding a pair of complex conjugate  solutions with
a very weakly decreasing real part ($\Omega_r=-1.67$ at
$\eta_i=1.78$ and $\Omega_r=-1.71$ at $\eta_i=10$).

Note that in all these cases the used  values of the ratios
$\eta_i\equiv L_n/L_{\sss T}$, $\eta_b\equiv L_n/L_{\sss B}$  in
fact imply a wide range of possible values for $L_n, \, L_{\sss T},
L_{\sss B}$. Since $\omega_r= \Omega_r \omega_{*e}\sim 1/L_n$, this
also implies a wide range of possible frequencies.

\section{Electric field}

\subsection{Acceleration of plasma particles}

The electrostatic instability discussed here implies an electric field varying in time and space, and having
very different components in the parallel and perpendicular directions. Such an electric field  can
accelerate plasma particles in both perpendicular and parallel directions with respect to the magnetic field vector. An
acceleration always exists in the presence of electrostatic perturbations, yet typically
it is sporadic and acts mainly on the small amount of particles from the far tail in the
distribution function. The critical value of such an electric field, above which the bulk
electron runaway effect takes place, in a fully ionized plasma is \citep{dr} $E_d=e
L_{ei}/(4 \pi \varepsilon_0 \lambda_d^2)$. Here, $L_{ei}=\log(\lambda_d/b)$ is the
Coulomb logarithm, $\lambda_d=\lambda_{de} \lambda_{di}/(\lambda_{de}^2+
\lambda_{di}^2)^{1/2}$ is the plasma Debye radius, $\lambda_{dj}=\vtj/\omega_{pj}$,
$\vtj, \omega_{pj}$ are, respectively,  the thermal velocity and the plasma frequency of
the $j$ species, and $b=[e^2/[12 \pi \varepsilon_0 (T_e+ T_i)]$ is the impact parameter
for electron-ion collisions. For the parameters used in the previous text  we have
$L_{ei}=19$, $\lambda_d=0.0005\;$m, and the Dreicer field is $0.11\;$V/m. Assuming the
parallel wave-length of about $100 (500)\;$km, the amplitude of the electrostatic
potential $\phi$ necessary to achieve the Dreicer value is about $1.8 (9)\;$KV. However,
in the perpendicular direction this same potential gives  the electric field $k_y \phi$
that is around $E_\bot=5.7 (29)\;$KV/m. For the parallel wave-length of about $1000\;$km we would
have $\phi=18\;$KV and consequently $E_\bot=57\;$KV/m. These estimates are for the number density
$n_0=10^{16}\;$m$^{-3}$.

\begin{figure}
\includegraphics[height=6cm, bb=15 15 283 226, clip=,width=.95\columnwidth]{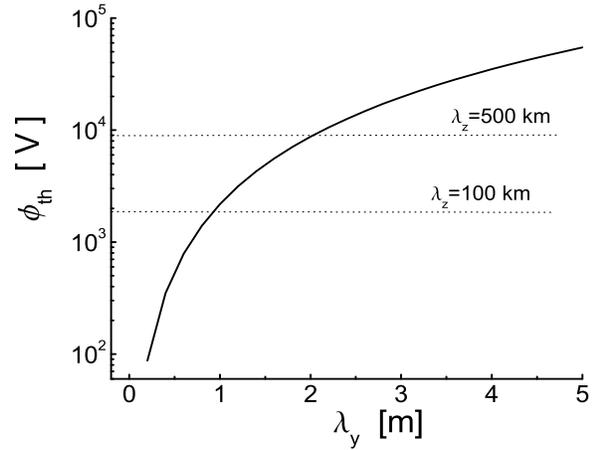}
\vspace*{3mm} \caption{The threshold value  of the potential $\phi$, Eq.~(\ref{h2}),  in terms of the perpendicular
wave-length, above which the stochastic heating takes place. Above the dotted line the parallel electric field exceeds
the Dreicer value, resulting in a simultaneous acceleration of particles. }\label{fig6}
\end{figure}

\begin{table*}
 \centering
 \begin{minipage}{70mm}
  \caption{Stochastic heating  for hydrogen  (and helium  in brackets) for several values of the
 perpendicular wavelength and  the wave amplitude $\phi$. The assumed starting temperature is $10^6$ K.}
  \begin{tabular}{@{}lllr@{}}
      \hline
      \hline
  $ \lambda_y$ [m]     &  $\phi $ [V]   &  $a$   &  $ T_{eff}$ [K]  \\
        \hline
   0.1  & $50$   & $2.3$ $(9.1) $  & $1.5  \cdot 10^6$ $  (2.6  \cdot 10^6)$      \\
  0.2  & $100$   & $1.1$ $ (4.6)  $  & $3 \cdot 10^6$ $  (3.5  \cdot 10^6)$      \\
   1   & $3000$  & $1.4$ $ (5.5) $  & $9 \cdot 10^7$ $  (1.2  \cdot 10^8) $      \\
    1  & $5000$  & $2.3$ $(9.1)$  & $1.5 \cdot 10^8 $ $ (2.6  \cdot 10^8)$      \\
     2  & $10000$  & $1.1$ $(4.6)$  & $3.1 \cdot 10^8 $ $ (3.5  \cdot 10^8)$      \\
\hline
 \hline
\end{tabular}
\end{minipage}
\end{table*}

Taking the number density one order of magnitude larger
$n_0=10^{17}\;$m$^{-3}$ yields  $E_d=1\;$V/m. The electric field
corresponding to this value in the parallel direction would (for the three
given parallel wavelengths)  in the perpendicular direction have the
magnitude of $54$, $270$, and $540\;$kV/m, respectively.

The time needed for the perturbations to achieve such values can be
estimated from the previously calculated growth-rates. Taking as an
example Fig.~3, the maximum growth-rate is $\gamma/\omega_{*e}\simeq
1$. Taking $L_n=10^3\;$m, $T_i=T_e=T_0=10^6\;$K, $B_0=3\cdot 10^{-2}\;$T,
for $\lambda_y=2\;$m we have $\omega_{*e}=9\;$Hz. Assuming some
small starting value of the electrostatic potential $\Psi$, the
growth time till it gets some value $\phi_1$ is $t_g\simeq
\log(\phi_1/\Psi)/\gamma$. Taking $e \Psi/(\kappa T_i)=0.01$ this
yields $\Psi=0.86\;$V.  Hence, the value $\phi_1=9\;$KV discussed above is
achieved within  $t_g\simeq 1\;$s.
Observe that for $L_n=10^4\;$m, this growth time becomes $10$ seconds.

We conclude that  the presented instability can yield the extremely large
values of the  electric field  reported in the observations
\citep{dav,zs} {\em within seconds}. The electric field generated  in such a way  may exceed
the Dreicer value, and consequently an acceleration of the bulk plasma may take
place. In the direction perpendicular to the magnetic field vector,  the particles are subject to the leading order
$\bmath{ E}\times \bmath{ B}$-drift $v_{\sss E}=E_\bot/B_0$, which is the same for both electrons and ions. Note that for $E_\bot=20\;$KV/m and
the earlier assumed magnetic field, it is of the order of $700\;$km/s. Hence, because of such short perpendicular wave-lengths, these are very  short-scale perpendicular plasma fluxes acting on the plasma as a whole.

The nature of the instability is that in the perpendicular direction (here along the $x$-axis)  it is localized in the  area of the maximum gradients e.g.\ around some value $x_0$. On the other hand, the perpendicular component of the perturbed electric field is in the $y$-direction (corresponding to the azimuthal direction in the cylindric geometry), so that due to the  $\bmath{ E}\times \bmath{ B}$-drift the plasma fluxes will be in the $x$-direction. Hence, the starting regular density profile in the $x$-direction will be modified and density condensations will appear one after another as one moves in the $y$-direction, left and right of the position $x_0$. In a realistic cylindric geometry and taking the mode behavior in the axial direction into account, this would imply the formation of braided density structures twisted along the cylinder.  However, in the limit of large potential amplitudes the initial plasma configuration can simply be destroyed.

In the parallel direction  the  velocity \citep{bit,v3}  is $v_{jz}(t)\sim E_z/[m_j(k_z v_0- \omega_r)]$. Hence,  the acceleration is proportional to $m_j^{-1}$  and acts mainly on electrons. It is selective in the sense that  particularly strong acceleration is experienced by resonant electrons having the starting velocity $\bmath{ v_0}$ satisfying the condition $v_{0z}=\omega_r/k_z$. More details on that issue may be found  in \citet{fh} and \citet{v3}.

\subsection{Quasi-static purely growing instability}

The instability described  in the previous text  implies the
presence of  a range of values for the three parameters $L_n,
L_{\sss T}, L_{\sss B}$ for which $|\Omega_r|\ll \gamma$. This is
due to the demonstrated  change of the mode direction, during which
the dispersion lines intersect with the $\eta_b$-axis so that $\omega_r=0$. In Figs.~3 and 4
this is around $\eta_b\simeq 0.25$. The corresponding growth-rate
for the two cases is $\gamma\simeq 0.6$ and $0.9$, respectively.

This implies an almost purely growing, quasi-static electric field, yet
spatially varying and with its  amplitude determined by the the two
mode numbers $k_{y,z}$. The above described acceleration will remain similar, yet the important difference is that
the mode is practically non-propagating and the spatial variation of the acceleration will become much more pronounced and the mentioned braiding more effective.

\subsection{Stochastic heating}

The polarization drift, i.e.\ the third term  in Eq.~(\ref{vel}),  starts to play an important role for a large enough
wave amplitude. In this case, the motion of a particle  becomes stochastic and consequently heating takes place.
Details of this process  can be found in \citet{bel}. It turns out that for a large enough wave amplitude the standard
iterative procedure,  which is behind Eq. (\ref{vel}), is not valid any more, and the same holds for  the particle
representation by its gyro-center. Instead, one is supposed to describe the actual particle motion by writing the
particle momentum equation for its motion in the wave-field.
 This has been described in detail and
experimentally verified  in \citet{san} and \citet{san2}. It is
shown that the stochastic heating takes place provided that
\be
a= k_y^2 \rho_i^2 \cdot \frac{e \phi }{T_{i0}} \geq 1. \label{h2} \ee
Here, $\rho_i=\vti/\Omega_i$. The condition (\ref{h2}) in fact implies  that in this regime the ion displacement due to the polarization drift
is comparable to the perpendicular wavelength. This is because $\bmath{ v_{pi}}= ( \bmath{ e_z}\times \partial \bmath{ v_{i\bot}}/\partial t)/\Omega_i$, and $\bmath{ v_{i\bot}}$ is the leading order
$\bmath{ E}\times \bmath{ B}$-drift,
so that \citep{san} $v_{pi}\sim a \omega/k_y$ and the perpendicular displacement due to the polarization drift
is $\delta=v_{pi}/\omega=a/k_y$.
 For these reasons the mentioned gyro-center representation fails and the field magnitude is to be calculated at the actual position of the particle.
 Another important feature
is that $\bmath{ v_{pi}} \sim \bmath{ k_y}$, hence the stochastic heating is due to the electrostatic property of the wave.
 Although the resulting particle motion is deterministic, as explained in \citet{bel}  the practical consequence of the described mechanism on the particle distribution function
is the same as in an ordinary  heating.

The maximum achieved stochastic temperature is \citep{san,san2}
 \be
  T_{max}= \frac{m}{3} \left(k_y^2 \rho_i^2 e \phi/T_{i0} + 1.9\right)^2\frac{\Omega_i^2}{k_y^2}. \label{vm} \ee

The application of this effect to the heating of the solar corona by the ordinary density gradient driven drift wave has been performed in our recent publications \citet{v2} and \cite{v3}. It is shown that the
ions are more efficiently heated than electrons, and the heavier ions are  heated better
than light ions provided that $k_y^4\rho_i^4 (e \phi/T_{i0})^2 >1.9$. The nature of the
heating is such that it acts  mainly in the perpendicular direction, and it can also
describe some other features of the coronal heating.

The threshold potential (\ref{h2}) is presented in Fig.~6 in terms of $\lambda_y$, for the same parameters as earlier
in the text. For the values above the curve, the stochastic heating takes place. On the other hand, as discussed
earlier in the text,  for large enough $k_z \phi$ the electric field exceeds the Dreicer value and the particle
acceleration is in action too.   The horizontal lines in Fig.~6 give $k_z\phi=E_d$ for the cases $\lambda_z=100\;$km and $500\;$km. Hence, for the values of $\phi$ above the both full and dotted lines, the plasma is subject to simultaneous heating
of ions (in the perpendicular direction), and an acceleration of bulk electrons in the parallel direction.

The stochastic temperature (\ref{vm}) is sensitive to the perpendicular wavelength and the magnitude of the background magnetic field. Using the parameters
from the previous text and the starting temperature of $1\;$MK one can easily obtain a temperature several orders of magnitude above the starting value, and this already for a very small amplitude of the potential $\phi$. The heating  presented in Table~1 is for rather moderate values of the perturbed potential, i.e., for the left (lower) part of the curve in Fig.~6. From Table~1, one concludes that at very short perpendicular wave-lengths, the stochastic heating is in action already at very small amplitudes of the wave potential. Also clear from Table~1 is the remarkable fact that for the same wave amplitude the heavier ions (helium) are heated more efficiently.

The stochastic heating implies the condition (\ref{h2}) satisfied, regardless of the specific values of the two separate parts in that expression, while the instability analysis and the  growth-time from Sec.~2 imply $k_y \rho_s\equiv  k_y \rho_i < 1$. For that reason we are formally allowed to estimate the growth time in Table~1 for $\lambda_y\geq 2\;$m only, and this is of the order of a second, as shown in Sec.~3.1. For this wave-length the maximum energy release rate per unit  volume for the density $n_0=10^{16}\;$m$^{-3}$ becomes $3 n_0 T_{max}/(2 t_g)=65\;$J/(m$^3$s), and that is several orders of magnitude above the value accepted as necessary for a sustained heating in coronal loops and also  well within the range of the total amount of the energy released in nano-flares. Some additional properties of the stochastic  heating, presented  in \citet{v2} and \citet{v3} for the ordinary drift wave, remain  valid for the present case as well.

\section*{Summary}
The heating of the solar corona  by waves, and the propagation of waves in the solar  environment  has been the subject of numerous studies in the past. Typically the wave heating models remain within the widely used MHD, e.g. \citet{pek, suz}, yet it is obvious that in such an approach a lot of physics remains out of scope as may be seen in \citet{v1, pw}.
  The analysis performed in the present paper is aimed at showing even further some novel phenomena that follow from a multi-component description. It is  based on the well established theory of the low frequency phenomena in magnetized and inhomogeneous plasmas. It is crucially a multi-component plasma description. In the same time, it represents a step forward  in the recently published modeling of the heating of the solar corona \citep{v2,v3}.  The most important consequence of the here described fast growing mode is the  electric field, with its drastically different scales in the parallel and perpendicular directions, that  cannot be predicted within the widely used MHD theory. This electric field should be responsible for the acceleration of plasma particles and their simultaneous heating.  The nature of the instability implies large scales in the direction of the  magnetic field vector. In a realistic cylindric configuration the mode is weakly twisted around a magnetic loop, and such is the scene where the predicted acceleration and heating takes place. The essential difference between the instability presented here  and the drift wave instability from   \citet{v2,v3} is that the former can be well described within the multi-component fluid theory, while the  latter is a purely kinetic effect.

The gradient driven oscillations presented here are of relatively high frequency and therefore presently difficult to
detect directly. Yet, they imply the presence of the electric fields  that  can be observed in coronal spectra due to
line broadening and shifts, that are  measured  and  presented in \citet{dav} and \citet{zs}, and attributed to plasma
waves and instabilities, in particular to the lower-hybrid-drift  and whistler instabilities. We show here that they
can be described in terms of the  gradient-driven drift instabilities. The possible consequences of the presence of
gradient driven oscillations (the temperature anisotropy and a  stronger heating of heaver ions) are well  documented
in the measurements of the spectral line  widths of heavy ions \citet{cr3}. The mentioned strong $\bmath{E}\times
\bmath{ B}$-plasma drifts should also be detectable as Doppler shifts in the spectra.

One can with certainty claim that, at least in the starting stages of the presented gradient-driven instability, these
processes are reasonably accurately described within the given  model  that can potentially be used for  the
description of the particle acceleration and the heating of the solar corona. However,  there exists  a number of
phenomena that  will mostly negatively affect the proposed effects of heating and acceleration, like  the energy and
particle  diffusion and collisions, coupling to the Alfv\'{e}n wave, non-linearity, etc.  Large wave amplitudes and the
corresponding  stronger heating  will necessarily modify the starting plasma configuration, primarily the temperature
gradient that is essential for the growth, and  the density gradient too. Consequently, the values obtained here
analytically  may  be far from accurate.   In particular, for large values of $e \phi/(\kappa T_i)$, various new
effects must be included, and  the model must be considerably improved in order to give accurate estimates in this
limit. Though, the effective temperature presented in Table~1 goes over  300 million K, that is  far above the
temperatures that are needed. Hence, reducing the potential [i.e. the factor $e \phi/(\kappa T_i)$] by two order of
magnitude will still give desired temperature of a few million K. Such a  reduction of $\phi$ will give lower values of
the electric field $k_y \phi$, yet  this can be compensated by taking larger values of $k_y$. For the given plasma
parameters,  the plasma Debye radius is around 0.5 mm,  so that going to very short perpendicular wavelengths, even
below the values from Table~1 is justified.  Nevertheless, it is fair to say that a realistic (non-linear) development
of the instability  and of all the consequences that follow from it,  can be described only numerically in codes that
allow for a simultaneous change of the driving force (i.e., the  inhomogeneous plasma background) in the process of the
development of the instability.

\section*{Acknowledgments}

These results were obtained in the framework of the projects
GOA/2009-009 (K.U.Leuven), G.0304.07 (FWO-Vlaanderen) and
C~90347 (ESA Prodex 9). Financial support by the European Commission through the SOLAIRE
Network (MTRN-CT-2006-035484) is gratefully acknowledged.

\bsp

\label{lastpage}

\end{document}